\pdfoutput=1
\documentclass[aps,prd,onecolumn,twoside,notitlepage,preprintnumbers,showpacs,showkeys,superscriptaddress,byrevtex,floatfix,nofootinbib,amsmath,amssymb]
{revtex4}
\usepackage{array,dsfont,graphicx,mathtools,placeins,verbatim,tensor}
\usepackage[bookmarksnumbered,bookmarksopen=true,bookmarksopenlevel=1,pdfstartview=FitH]{hyperref}
\usepackage[all]{hypcap}

\newcolumntype{M}[1]{>{$}{#1}<{$}}

\def\0{{\sst{(0)}}}
\def\1{{\sst{(1)}}}
\def\2{{\sst{(2)}}}
\def\3{{\sst{(3)}}}
\def\4{{\sst{(4)}}}
\def\5{{\sst{(5)}}}
\def\6{{\sst{(6)}}}
\def\7{{\sst{(7)}}}

\def\nn{\nonumber} \def\bd{\begin{document}} \def\ed{\end{document}}
\def\ds{\documentstyle} \let\fr=\frac \let\bl=\bigl \let\br=\bigr
\let\Br=\Bigr \let\Bl=\Bigl 
\let\bm=\bibitem
\let\na=\nabla
\let\pa=\partial \let\ov=\overline 
\newcommand{\be}{\begin{equation}} 
\newcommand{\ee}{\end{equation}} 
\def\ba{\begin{array}}
\def\ea{\end{array}}
\def\ft#1#2{{\textstyle{{\scriptstyle #1}\over {\scriptstyle #2}}}}
\def\fft#1#2{{#1 \over #2}}
\def\del{\partial}
\def\sst#1{{\scriptscriptstyle #1}}
\def\oneone{\rlap 1\mkern4mu{\rm l}}
\def\ie{{\it i.e.\ }}
\def\via{{\it via}}
\def\semi{{\ltimes}}
\def\v{{\cal V}}

\newcommand{\bp}{\bullet}
\newcommand{\sfx}{\textsf{x}}
\newcommand{\sfo}{\textsf{o}}
\newcommand{\alg}[1]{\ensuremath{\mathfrak{#1}}}
\newcommand{\rep}[1]{\ensuremath{\mathbf{#1}}}
\newcommand{\fld}[1]{\ensuremath{\mathds{#1}}}
\newcommand{\rng}[1]{\ensuremath{\mathds{#1}}}
\newcommand{\SUSY}{\ensuremath{\mathcal{N}}}

\newcommand{\ho}[1]{$\, ^{#1}$}
\newcommand{\hoch}[1]{$\, ^{#1}$}
\newcommand{\bea}{\begin{eqnarray}} 
\newcommand{\eea}{\end{eqnarray}} 
\newcommand{\ra}{\rightarrow}
\newcommand{\lra}{\longrightarrow}
\newcommand{\Lra}{\Leftrightarrow}
\newcommand{\tr}{{\rm tr} }
\newcommand{\Tr}{{\rm Tr} } 

\DeclareMathOperator{\Span}{span}

\begin{document}

\title{Four curious  supergravities}
\author{M. J. Duff}
\email[]{m.duff@imperial.ac.uk}
\affiliation{Theoretical Physics, Blackett Laboratory, Imperial College London, London SW7 2AZ, United Kingdom}
\author{S. Ferrara}
\email[]{sergio.ferrara@cern.ch}
\affiliation{Physics Department, Theory Unit, CERN, CH 1211, Geneva 23, Switzerland}
\affiliation{INFN - Laboratori Nazionali di Frascati, Via Enrico Fermi 40, 00044 Frascati, Italy}
\affiliation{Department of Physics and Astronomy, University of California, Los Angeles, CA USA}
\date{\today}

\begin{abstract}

We consider four supergravities with $16+16$, $32+32$, $64+64$, $128+128$ degrees of freedom displaying some curious properties: (1) They exhibit minimal supersymmetry (${\cal N}=1, 2, 2, 1$) but maximal rank ($r=7, 6, 4, 0$) of the scalar coset in $D=4, 5, 7, 11$.  (2) They couple naturally to supermembranes and admit these membranes as solutions.  (3) Although the $D=4, 5, 7$ supergravities follow from truncating the maximally supersymmetric ones, there nevertheless exist M-theory compactifications with $G_2$, $SU(3)$, $SU(2)$ holonomy having these supergravities as their massless sectors.  (4) They reduce to $\SUSY=1, 2, 4, 8$ theories all with maximum rank 7 in $D=4$ which (5) correspond to $0, 1, 3, 7$ lines of the Fano plane and hence admit a division algebra \fld{(R,C,H,O)} interpretation consistent with the black-hole/qubit correspondence, (6)  are generalized self-mirror and hence (7) have vanishing on-shell trace anomaly.

\end{abstract}
\begin{flushright}
\hfill{Imperial/TP/2010/mjd/4}\\
\hfill{CERN-PH-TH/2010-220}\\
\end{flushright}
\pacs{11.25.-w}
\keywords{keywords}

\maketitle
\tableofcontents
\newpage
\FloatBarrier
\section{Introduction}
\label{Introduction}

In the early eighties Green and Schwarz \cite{Green:1983wt} showed that spacetime supersymmetry allows classical superstrings moving in spacetime dimensions $3,4,6$ and $10$, with the $D=10$ case being anomaly-free at the quantum level.  However, following the branescan \cite{Achucarro:1987nc} of  \autoref{kappascan}, which pinpointed those twelve $(p,D)$ slots consistent with kappa-symmetric Green-Schwarz  type actions\footnote{ It will be sufficient for our purposes to focus on this old branescan with just scalars and spinors on the worldvolume, as opposed to the new one \cite{Duff:1992hu,Duff:1994an} with vectors (D-branes) and tensors (M5-branes) also.}, it was realized that these $1$-branes in $D=3,4,6$ and $10$ should now be viewed as the endpoints of four sequences of $p$-branes. Moving diagonally down the brane-scan corresponds to a simultaneous dimensional reduction of spacetime and worldvolume \cite{Duff:1987bx}.  In particular, supermembranes exist in $D= 4,5,7,11$ with minimal supersymmetry ${\cal N}=1, 2, 2, 1$, respectively.

\begin{table}[h!]
\begin{center}
\begin{tabular}{ccccccccccccccc}
~&$D\uparrow$&&&&&&&&&&&~\\
~&11&.&&&o&&&&&&&&~\\
~&10&.&&o&&&&o&&&&&~\\
~&9&.&&&&&o&&&&&&~\\
~&8&.&&&&o&&&&&&&~\\
~&7&.&&&o&&&&&&&&~\\
~&6&.&&o&&o&&&&&&&~\\
~&5&.&&&o&&&&&&&&~\\
~&4&.&&o&o&&&&&&&&~\\
~&3&.&&o&&&&&&&&&~\\
~&2&.&&&&&&&&&&&~\\
~&1&.&~&~&~&~&~&~&~&~&~&~&~\\
~&0&.&.&.&.&.&.&.\\
~&~&0&1&2&3&4&5&6&$p+1\rightarrow$
\end{tabular}
\end{center}
\caption{ $p$-branes described by Green-Schwarz actions}
\label{kappascan}
\end{table}

In $D=11$, it is known that the membrane \cite{Bergshoeff:1987cm} couples to the $D=11$ supergravity background $(g_{MN};  \psi_{M};  A_{MNP})$, and that this supergravity in turn admits the membrane as a solution \cite{Duff:1990xz}. Yet little attention has been paid to the corresponding supergravities that couple to membranes in $D=4, 5$ and $7$ (or strings in $D=3, 4$ and $6$). This hitherto lack of interest\footnote{Exceptions may be found in \cite{Bergshoeff:1985su,Ovrut:1997ur,Baez:2009xt,Baez:2010ye} and in John Baez's blog http://golem.ph.utexas.edu/category/2010/03/division\_algebras\_and\_supersym.html.} in these minimal supergravity theories  is no doubt due to the perception that they describe the low-energy limit of  non-critical quantum inconsistent string or M theories. 
 
In this paper, however, we consider those supergravities  that emerge as the massless sectors of compactifications of M-theory on manifolds/orbifolds  $X^4$, $X^6$ and $X^7$  with reduced holonony $SU(2)$, $SU(3)$ and $G_2$  and with special betti numbers:
\[
X^4: (1, 0, 6, 0, 1)
\]
\[
X^6:(1, 0, 3, 8, 3, 0, 1)
\]
\begin{equation}
X^7: (1, 0, 0, 7, 7, 0, 0, 1)
\end{equation}
respectively.  This means that the resulting theories in $D=7, 5, 4$ with ${\cal N}=2, 2, 1$ are just as interesting as their  counterpart in $ D=11$ with ${\cal N}=1$.  Denoting the fields that result from the $D=11$ (metric; gravitino; 3-form) by $(g_{\mu\nu}, {\cal A}_{\mu}, {\cal A};  \psi_{\mu}, \chi; A_{\mu\nu\rho}, A_{\mu\nu}, A_{\mu}, A)$, and their numbers of degrees of freedom by $f$, we show in section \ref{minimal} that the four supergravities are:\begin{itemize}
\item
$D=11$: \SUSY=1 graviton, $f=128+128$
\[
(g_{\mu\nu}; \psi_{\mu}; A_{\mu\nu\rho})
\]
 \item
$ D=7$: \SUSY=2 graviton +3 vector,  $f=40+40 +3(8+8)=64+64$
 \[
(g_{\mu\nu}, {\cal A};  2\psi_{\mu}, 2\chi; A_{\mu\nu\rho}, 3A_{\mu} )+3(3{\cal A}; 2\chi; A_{\mu} )
\]
with rank 4 scalar coset
\[
\frac{G}{H}=SO(1,1) \times  \frac{SL(4,R)}{SO(4)}
\]
\item
$D=5$: \SUSY=2 graviton + 2 vector  + 3 hyper + 1 3-form, $f=8+8 +2(4+4)+3(4+4)+(4+4)=32+32$
\[
(g_{\mu\nu}; 2\psi_{\mu}; A_{\mu})+2({\cal A}; 2\chi; A_{\mu})+3(2{\cal A}; 2\chi; 2A)+ ({\cal A}; 2\chi; A_{\mu\nu\rho}, 2A)
\]
with rank 6 scalar coset
\[
\frac{G}{H}=SO(1,1)^3 \times  \frac{SO(3,4)}{SO(3) \times SO(4)} \ltimes R^2
\]

\item
$D=4$:  \SUSY=1 graviton +7 WZ,  $f=2+2 +7(2+2)=16+16$
\[
(g_{\mu\nu}; \psi_{\mu}; A_{\mu\nu\rho})+7({\cal A}; \chi; A)
\]
with rank 7 scalar coset
\[
\frac{G}{H}=\frac{SL(2)^7}{SO(2)^7} 
\]
The U-duality is only $SL(2)^6 \times SO(1,1) \ltimes R$, however, because of the coupling of the scalars to $A_{\mu\nu\rho}$.
\end{itemize}

These theories may also be derived by compactifying  M-theory on $T^4$, $T^6$ and $T^7$ and truncating the massless sectors so as to obtain minimal supersymmetry ($\SUSY=2,2$ and $1$ respectively) while preserving  maximum rank of the scalar coset  ($r=4,6$ and $7$ respectively).  They exhibit several other remarkable properties. For example, they admit  membranes as elementary (electric) solutions, by virtue of the universal presence of a 3-form $A_{\mu\nu\rho}$ and by virtue of the correct dilaton exponent which follows from the maximum rank condition. We therefore expect that they will be compatible with the superspace constraints enforced by kappa symmetry on the worldvolume of the Green-Schwarz membranes \cite{Bergshoeff:1987cm}, but we do not address this problem here.

It will be  important  for our purposes to distinguish between the Lagrangians obtained directly in these two ways and the conventional supergravity Lagrangians obtained after dualization of the $p$-forms. For example, the latter has no $A_{\mu\nu\rho}$ field in $D=4$ and so not only has a different symmetry, namely $SL(2)^7$ as opposed to the $SL(2)^6 \times SO(1,1) \ltimes R$ of the former, but also admits no electric membrane (domain wall) solution.

As described in section \ref{D4}, these four theories may be further reduced  to $\SUSY =1, 2, 4$ and $8$ theories all with maximum rank $r=7$ in $D=4$, corresponding to compactification on $X^7$ with independent betti numbers
\begin{equation}
(b_0,b_1,b_2,b_3)=(1,{\cal N}-1,3{\cal N}-3, 4{\cal N}+3)
\end{equation}
Compactifications with such betti numbers may indeed be found in \cite{Cremmer:1978km} for ${\cal N}=8$, in \cite{Ferrara:1989nm,Sen:1995ff} for ${\cal N}=4$ and in \cite{Joyce:1996a,Joyce:1996b,Gaberdiel:2004vx} for ${\cal N}=2$ and ${\cal N}=1$. We show that the corresponding supergravities before dualization are:
\begin{itemize}
\item{$T^7: (1, 7, 21, 35)$}

\SUSY=8 graviton, $f=128+128$,
\[
(g_{\mu\nu}, 7{\cal A}_{\mu}, 28 {\cal A}; 8\psi_{\mu}, 56\chi; A_{\mu\nu\rho}, 7A_{\mu\nu},21 A_{\mu},35 A)
\]
with rank 7 scalar coset
\[
\frac{G}{H}=SO(1,1) \times \frac{SL(7,R)}{SO(7)} \ltimes R^{35} 
\]

 \item{$X^4 \times T^3: (1,3,9,19)$}
 
\SUSY=4 graviton + 3 vector + 3 2-form,  $f=16+16 +3(8+8)+3(8+8)=64+64$, 
 \[
(g_{\mu\nu}, 3{\cal A}_{\mu}, {\cal A};  4\psi_{\mu}, 4\chi; A_{\mu\nu\rho}, 3A_{\mu}, A )+3(3{\cal A}; 4\chi; A_{\mu}, 3A)
+3(2{\cal A}; 4\chi; A_{\mu\nu}, A_{\mu}, 3A)
\]
with rank 7 scalar coset
\[
\frac{G}{H}=\frac{SL(2,R)}{SO(2)} \times \frac{SO(3,6)}{SO(3) \times SO(6)} \times SO(1,1) \times \frac{SL(3)}{SO(3)} \ltimes R^{9}
\]
\item{$X^6 \times S^1:(1, 1, 3,11)$}

\SUSY=2 graviton + 3 vector  + 3 hyper +1 linear, $ f=4+4 +3(4+4)+3(4+4)+(4+4)=32+32$, 
\[
(g_{\mu\nu}, {\cal A}_{\mu}; 2\psi_{\mu}; A_{\mu\nu\rho})+3({\cal A}; 2\chi; A_{\mu}, A)+3(2{\cal A}; 2\chi; 2A)+ ({\cal A}; 2\chi; A_{\mu\nu}, 2A)
\]
with rank 7 scalar coset
\[
\frac{G}{H}=SO(1,1) \times \frac{SL(2,R)^3}{SO(2)^3} \times \frac{SO(3,4)}{SO(3) \times SO(4)} \ltimes R^2
\]
\item{$X^7: (1, 0, 0, 7)$}

\SUSY=1 graviton +7 WZ,  $f=2+2 +7(2+2)=16+16$,  
\[
(g_{\mu\nu}; \psi_{\mu}; A_{\mu\nu\rho})+7({\cal A}; \chi; A)
\]
with rank 7 scalar coset
\[
\frac{G}{H}=\frac{SL(2)^7}{SO(2)^7}
\]

\end{itemize}

Interestingly enough, the cases ${\cal N}=8$,  ${\cal N}=4$ and  ${\cal N}=2$ (albeit without the three hyper and one linear multiplet \cite{D'Auria:2004yi}) have already made an appearance in the context of the {\it black-hole/qubit correspondence} \cite{Duff:2006uz,Kallosh:2006zs,Levay:2006kf},  where their $56, 24, 8$ black hole charges correspond to $7, 3, 1$ lines of the Fano plane \cite{Duff:2006ue,Levay:2006pt,Borsten:2008wd}. In particular, the ${\cal N}=2$ supergravity is just the STU model  whose black holes have a Bekenstein-Hawking entropy given by Cayley's hyperdeterminant, the same quantity that describes the entanglement of three qubits.  The  $7, 3, 1$ lines of the Fano plane in turn provide the multiplication table of  the imaginary octonion, quaternion, and complex numbers respectively.  The fourth ${\cal N}=1$ supergravity completes the set with $0$ lines, corresponding to the reals. 

Earlier work on the branescan gave an $\mathds{O,H,C,R}$ division algebra interpretation to the four sequences appearing in  \autoref{kappascan}, which have $8+8, 4+4, 2+2, 1+1$ worldvolume degrees of freedom. See \cite{Kugo:1982bn,Evans:1987tm,Duff:1987qa}  and references therein. Since our supergravites are obtained by compactification, however, the corresponding membranes all have  $8+8$.

 Furthermore, we recall that in \cite{Duff:2010ss} we defined a generalized mirror symmetry
\begin{equation}
(b_0,  b_1, b_2, b_3) \rightarrow (b_0, b_1,  b_2 -\rho/2, b_3+\rho/2)
\label{eq:mirror}
\end{equation}
under which
\begin{equation}
\rho \equiv 7b_0-5b_1+3b_2 +b_3
\end{equation}
changes sign
\begin{equation}
\rho \rightarrow -\rho
\end{equation}
Generalized self-mirror theories are defined to be those for which $\rho$ vanishes.  In the case of $G_2$ manifolds with $b_1=0$,  Joyce \cite{Joyce:1996a,Joyce:1996b} refers to $\rho=0$ as an ``axis of symmetry''.  For related work on mirror symmetry and Joyce-manifiolds, see \cite{Shatashvili:1994zw,Acharya:1997rh,Gaberdiel:2004vx}.

Moreover the quantity $\rho$ also shows up  in the on-shell Weyl anomaly \cite{Duff:1977ay,Duff:1993wm},  before dualization \cite{Duff:1980qv},  which is given by
\begin{equation}
g_{\mu\nu}<T^{\mu\nu}>=A \frac{1}{32\pi^2}R^{*}{}^{\mu\nu\rho\sigma}R^{*}{}_{\mu\nu\rho\sigma}
\end{equation}
where
\begin{equation}
A=-\frac{1}{24}\rho
\end{equation}
Since our four curious supergravities  all have $\rho=0$, they are self-mirror in the above sense  and hence have vanishing Weyl anomaly.

Finally we note that a spacelike reduction gives the four Type IIA supergravities that couple to superstrings in $D=3, 4, 6, 10$. They yield ${\cal N}=16,8,4,2$ supergravities in $D=3$. While a timelike reduction from $D=4$ to $D=3$ yields after dualization the four cosets that play a role in the four-way entanglement of eight qubits \cite{Borsten:2008wd,Levay:2010ua,Borsten:2010db,Borsten:2010ab, Gibbs:2010aa}, namely $E_{8(8)}/SO^*(16)$,  $SO(8,8)/SO(4,4)^2$, $SO(4,4)^2/SO(2,2)^4$, $SO(2,2)^4/SO(1,1)^8$.

\section{Minimal supergravities in $D=4,5,7,11$}
\label{minimal}
\FloatBarrier
\subsection{Compactifications}

\begin{table}[h!]
$\begin{array}{lrrrrrrrrrr}
Field &f& \\
&&&&\\
g_{MN}&44&1\\
\psi_{M}&128&1\\
A_{MNP}&84&1\\
&&&&&\\
total~f&&256\\
\end{array}$
\caption{ D=11 fields}
\label{X0}
\end{table}

To derive the $D=4,5,7,11$ theories we begin  with  compactification on generic manifolds, tori and manifolds of special holonomy as shown in \autoref{X0},  \autoref{X4},  \autoref{X6}  and \autoref{X7}

\begin{table}[h!]
$\begin{array}{llrrrrrrrrrr}
&Field &f&generic& torus &special& \\
&&&&&&&\\
\bigskip
&&&&\\
g_{MN}&g_{\mu\nu}&14&d_0&1&1\\
~&{\cal A}_{\mu}&5&d_1&4&0\\
~&{\cal A}&1&-8d_0+3d_2&10&10\\
\psi_{M}&\psi_{\mu}&16&2d_0+d_1/2&4&2\\
~&\chi&4&-4d_0+2d_1 +2d_2&16&8\\
A_{MNP}&A_{\mu\nu\rho}&10&d_0&1&1\\
~&A_{\mu\nu}&10&d_1&4&0\\
~&A_{\mu}&5&d_2&6&6\\
~&A&1&d_1&4&0\\
&&&&&\\
total~f&&&16(2d_0+2d_1+d_2)&256&128&\\
&&&&&\\
\chi(X^4)&&&2d_0-2d_1+d_2&0&8
\end{array}$
\caption{ Compactify to $D=7$ on $X^4$ with betti numbers: generic $(d_0=d_4=1; d_1=d_3; d_2)$; torus $(1; 4; 6)$ and $SU(2)$ holonomy  $(1; 0; 6)$.}
\label{X4}
\end{table}

\begin{table}[h!]
$\begin{array}{llrrrrrrrrrr}
&Field &f&generic& torus &special& \\
&&&&&&&\\
\bigskip
&&&&\\
g_{MN}&g_{\mu\nu}&5&c_0&1&1\\
~&{\cal A}_{\mu}&3&c_1&6&0\\
~&{\cal A}&1&-2c_0-2c_1+c_2+c_3 &21&9\\
\psi_{M}&\psi_{\mu}&4&2c_0+c_1&8&2\\
~&\chi&2&-2c_0+2c_2+c_3&48&12\\
A_{MNP}&A_{\mu\nu\rho}&1&c_0&1&1\\
~&A_{\mu\nu}&3&c_1&6&0\\
~&A_{\mu}&3&c_2&15&3\\
~&A&1&c_3&20&8\\
&&&&&&&\\
total~f&&&4(2c_0+2c_1+2c_2+c_3)&256&64&\\
&&&&&&&\\
\chi(X^6)&&&2c_0-2c_1+2c_2-c_3&0&0&\\
&&&&&\\
&&&\end{array}$
\caption{ Compactify to $D=5$ on $X^6$ with betti numbers: generic $(c_0=c_6=1; c_1=c_5; c_2=c_4; c_3)$, torus  $(1; 6; 15; 20)$ and $SU(3)$ holonomy $(1; 0; 3; 8)$}
\label{X6}
\end{table}

\begin{table}[h!]
$\begin{array}{llrrrrrrrrrr}
&Field &f&generic& torus &special& \\
&&&&&&&\\
\bigskip
&&&&\\
g_{MN}&g_{\mu\nu}&2&b_0&1&1\\
~&{\cal A}_{\mu}&2&b_1&7&0\\  
~&{\cal A}&1&-b_1+b_3 &28&7\\
\psi_{M}&\psi_{\mu}&2&b_0+b_1&8&1\\
~&\chi&2&b_2+b_3&56&7\\
A_{MNP}&A_{\mu\nu\rho}&0&b_0&1&1\\
~&A_{\mu\nu}&1&b_1&7&0\\
~&A_{\mu}&2&b_2&21&0\\
~&A&1&b_3&35&7\\
&&&&&\\
total~f&&&4(b_0+b_1+b_2+b_3)&256&32&\\
&&&&&\\
\rho(X^7)&&&7b_0-5b_1+3b_2-b_3&0&0&\\
\end{array}$
\caption{ Compactify to $D=4$ on $X^7$ with betti numbers: generic $(b_0=b_7=1; b_1=b_6; b_2=b_5, b_3=b_4)$, torus $(1; 7; 21; 35)$ and $G_2$ holonomy $(1; 0; 0; 7)$}
\label{X7}
\end{table}

\FloatBarrier
 \subsection{Supermultiplets}

 Here we group the individual fields  into supermutiplets:

 \begin{table}[h!]
$\begin{array}{llrrrrrrrrrr}
N=1~ &  multiplet&f& \\
 &&\\
&&&&&&&\\
\bigskip
graviton&(g_{MN}; \psi_{M}; A_{MNP})&128+128&\\
\end{array}$
\caption{ The $D=11$ multiplet in the minimal  $\mathcal{N}=1$ basis}
\label{Y0}
\end{table}

\begin{table}[h!]
$\begin{array}{llrrrrrrrrrr}
N=2~  & multiplet&f& N=2d_0+d_1/2 &  N=4 &  N=2 & \\
&&&&&&&\\
\bigskip
graviton&(g_{\mu\nu},  {\cal A}; 2\psi_{\mu}, 2\chi; A_{\mu\nu\rho}, 3A_{\mu})&40+40&d_0&1&1\\
gravitino&(4{\cal A}_{\mu}; 2\psi_{\mu}, 8\chi; 4A_{\mu\nu}, 4A) &64+64&d_1/4&1&0 \\
&&&&&&&\\
vector&(3{\cal A}; 2\chi; A_{\mu})&8+8&-3d_0+d_2&3&3\\
\end{array}$
\caption{ The $D=7$ multiplets in the minimal $\mathcal{N}=2$ basis}
\label{Y4}
\end{table}

\begin{table}[h!]
$\begin{array}{llrrrrrrrrrr}
N=2~ & multiplet &f&N=2c_0+c_1 &  N=8 &  N=2 & \\
&&&&&&&\\
\bigskip
graviton&(g_{\mu\nu}; 2\psi_{\mu}; A_{\mu})&8+8&c_0&1&1\\
gravitino&(2{\cal A}_{\mu}; 2\psi_{\mu}, 2\chi; 2A_{\mu}) &12+12&c_1/2&3&0 \\
&&&&&&&\\
vector&({\cal A}; 2\chi; A_{\mu})&4+4&-c_0-c_1+c_2&8&2\\
&&&&&&&\\
hyper&(2{\cal A}; 2\chi; 2A)&4+4&-c_0-c_1/2+c_3/2&6&3\\
&&&&&&\\
2-form&(2\chi; A_{\mu\nu}, A)&4+4&c_1&6&0\\
&&&&&&\\
3-form&({\cal A}; 2\chi; A_{\mu\nu\rho}, 2A)&4+4&c_0&1&1
\end{array}$
\caption{  The $D=5$ multiplets in the minimal $\mathcal{N}=2$ basis}
\label{Y6}
\end{table}

\begin{table}[h!]
$\begin{array}{llrrrrrrrrrr}
N=1~& multiplet  &f&N=b_0+b_1 &  N=8 &  N=1 & \\
 &&&&&&&\\
\bigskip
graviton&(g_{\mu\nu};\psi_{\mu}; A_{\mu\nu\rho})&2+2&b_0&1&1\\
gravitino&({\cal A}_{\mu}; \psi_{\mu}) &2+2&b_1&7&0 \\
&&&&&&&\\
vector&(\chi; A_{\mu})&2+2&b_2&21&7\\
&&&&&&&\\
 WZ &({\cal A}; \chi; A)&2+2&-b_1+b_3&28&7\\
&&&&&&\\
linear&(\chi; A_{\mu\nu}, A)&2+2&b_1&1&1
\end{array}$
\caption{ The $D=4$ multiplets in the minimal $\mathcal{N}=1$ basis}
\label{Y7}
\end{table}

\FloatBarrier
\subsection{Lagrangians }
\label{lagrangians}

 The bosonic sector of the toroidally compactified $D=11$  supergravity prior to dualization may be found in \cite{Cremmer:1997ct,Cremmer:1998px}.  It will be useful to split the metric scalars  ${\cal A}$ into $\vec\phi$,  the $(11-D)$-vector of dilatonic scalar fields coming from the diagonal components of the internal metric, and the rest, which we continue to describe by the letter ${\cal A}$.  The original eleven-dimensional fields
$g_{\sst{MN}}$ and $A_{\sst{MNP}}$ will give then rise to the following
fields in $D$ dimensions,
\bea
g_{\sst{MN}} &\rightarrow & g_{\mu\nu} , \qquad \vec\phi,\qquad
{\cal A}_{\mu}^{i},\qquad {\cal A}^i{}_j  \nn\\
A_{MNP} &\rightarrow & A_{\mu\nu\rho} ,\qquad A_{\mu\nu k }, \qquad A_{\mu jk},
\qquad A_{ijk}
\label{dfields}
\eea
where the indices $i, j, k$ run over the $(11-D)$ internal toroidally-compactified dimensions. If we denote the rank $p+1$ field strengths of the rank $p$ potentials by a subscript $(p+1)$, the Lagrangian is
\bea
\frac{{\cal L}}{\sqrt{-g}} &=& R -\ft12 \, (\del\vec\phi)^2 -\ft1{48}\, e^{\vec a\cdot
\vec\phi}\, F_\4^2 -\ft{1}{12} \sum_i
e^{\vec a_i\cdot \vec\phi}\, (F_{\3i})^2
-\ft14\, \sum_{i<j} e^{\vec a_{ij}\cdot \vec\phi}\, (F_{\2ij})^2
\label{dgenlag}\\
&& -\ft14\, \sum_i e^{\vec b_i\cdot \vec\phi}\, ({\cal F}_\2^i)^2
-\ft12 \, \sum_{i<j<k} e^{\vec a_{ijk} \cdot\vec \phi}\,
(F_{\1ijk})^2 -\ft12\, \sum_{i<j} e^{\vec b_{ij}\cdot \vec\phi}\,
({\cal F}_\1^i{}_j)^2 + {\cal L}_{\sst{FFA}}\ \nn
\eea
where the ``dilaton vectors'' $\vec a$, $\vec a_i$, $\vec a_{ij}$,
$\vec a_{ijk}$,
$\vec b_i$, $\vec b_{ij}$ are constants that characterise the couplings of
the dilatonic scalars $\vec \phi$ to the various gauge fields \cite{Lu:1995yn}
\bea
{F}_4:&&\vec a = -\vec g\ \\
{F}_3:&&\vec a_i = \vec f_i -\vec g \ \\
{F}_2:&& \vec a_{ij} = \vec f_i + \vec f_j - \vec g\ \\
{F}_1:&&\vec a_{ijk} = \vec f_i + \vec f_j + \vec f_k -\vec g\\
{\cal F}_2:&&\vec b_i = -\vec f_i \ \\
{\cal F}_1:&& b_{ij} = -\vec f_i + \vec f_j\ 
\label{dilatonvec}\\
\eea
where the vectors $\vec g$ and $\vec f_i$ have $(11-D)$ components
in $D$ dimensions, and are given by
\bea
\vec g &=&3 (s_1, s_2, \ldots, s_{11-D})\ ,\nonumber\\
\vec f_i &=& \Big(\underbrace{0,0,\ldots, 0}, (10-i) s_i, s_{i+1},
s_{i+2}, \ldots, s_{11-D}\Big)\ \label{gfdef}
\eea
where $s_i = \sqrt{2/((10-i)(9-i))}$. Note that the $4$-dimensional
metric is related to the eleven-dimensional one by
\be
ds_{11}^2 = e^{\ft13 \vec g\cdot\vec\phi} \, ds_{\sst 4}^2 +
\sum_i e^{2\vec\gamma_i\cdot\vec\phi}\, (h^i)^2\ \label{met}
\ee
where 
\be
\vec \gamma_i=\ft16\vec g -\ft12\vec f_i
\ee 
and
\be
h^i=dz^i + {\cal A}^i + {\cal A}^i{}_j\, dz^j\ 
\ee
 In general, the field strengths appearing in the kinetic
terms are not simply the exterior derivatives of their associated
potentials, but have non-linear Kaluza-Klein modifications as well.
On the other hand the terms included in ${\cal L}_{\sst{FFA}}$, which
denotes the dimensional reduction of the $F_\4\wedge F_\4\wedge A_\3$
term in $D=11$, are best expressed purely in terms of the potentials
and their exterior derivatives.  The complete details may be found in \cite{Lu:1995yn}, where it is shown that the symmetry of the Lagrangian is
\be
GL(11-D,R) \ltimes R^q
\ee
with
\be
q=\frac{1}{6}(11-D)(10-D)(9-D)
\ee

 Our {\it minimal} Lagrangians in $D=7,5,4$ follow by appropriate truncations that nevertheless keep all the  $\vec\phi$. We shall not show these explicitly.

\FloatBarrier
\subsection{Membrane solutions}

According to \cite{Duff:1994an,Lu:1995yn,Lavrinenko:1996mp},  the existence of an elementary membrane solution in $D$ dimensions requires a metric,  3-form potential and dilaton described by the action 
\be
\frac{{\cal L}}{\sqrt{-g}} = R -\ft12 \, (\del\phi)^2 -\ft1{48}\, e^{a\phi}\, F_\4^2 
\label{membarn}
\ee
Moreover the dilaton coupling must be such that
\be
a^2=\frac{2(11-D)}{(D-2)}
\ee
or $a=0,  8/5, 4, 7$  in $D=11, 7, 5, 4.$  But if we start with 
\bea
\frac{{\cal L}}{\sqrt{-g}} &=& R -\ft12 \, (\del\vec\phi)^2 -\ft1{48}\, e^{\vec a \cdot
\vec\phi}\, F_\4^2 
\eea
and make the ansatz
\be
a \vec \phi= {\vec a \phi}
\ee
noting that
\be
 \vec a . \vec a= \frac{2(11-D)}{(D-2)}=a^2
 \ee
 then the two Lagrangians coincide. Note that this ansatz would not have worked had we failed to implement the maximum rank condition by omitting some of the components of $ \vec \phi $. 

\FloatBarrier
\subsection{Cosets}
The scalar cosets, before and after dualization are shown in Table \ref{cosets1} and \ref{cosets2}
\begin{table}[h!]
$\begin{array}{llllllllcccccc}
theory&charges&G/H&dim&rank&max~G/H&dim&rank\\
&&&&&\\
 D=11  & 32&  0 &0&0&\subset 0&0&0 \\
D=7&16&SO(1,1) \times SL(4,R)/SO(4)&10&4&\subset SL(5,R)/SO(5) &14&4\\
D=5&8&SO(1,1)^3 \times SO(4,3)/[SO(4)\times SO(3)]\ltimes R^{2}&17&6&\subset SO(1,1) \times SL(6,R)/SO(6) \ltimes R^{20}&41&6&\\
D=4&4&SL(2,R)^7/SO(2)^7 &14&7&\subset SO(1,1) \times SL(7,R)/SO(7) \ltimes R^{35}&63 &7\\
\end{array}$
\label{cosets1}
\caption{$D=4,5,7,11$ cosets before dualization}
\end{table}
\begin{table}[h!]
$\begin{array}{llllllllccccccccc}
theory&charges&G/H&dim&rank&max~G/H&&&&&&&&&dim&rank\\
&&&&&&&&&&&&\\
 D=11  & 32   &0&0&0&\subset 0&&&&&&&&&0&0 \\
D=7&16&SO(1,1) \times SL(4,R)/SO(4) &10&4&\subset SL(5,R)/SO(5)&&&&&&&&&14&4\\
D=5&8&SO(1,1)^2 \times SO(4,4)/SO(4)^2& 18&6&\subset E_{6(6)}/Usp(8)&&&&&&&&&42&6\\
D=4&4&SL(2,R)^7/SO(2)^7 &14&7&\subset E_{7(7)}/SU(8)&&&&&&&&&70 &7\\
\end{array}$
\label{cosets2}
\caption{$D=4,5,7,11$ cosets after dualization}
\end{table}

\newpage
\section{$\mathcal{N}=1,2,4,8$ in $D=4$}
\label{D4}
\FloatBarrier
\subsection{Betti numbers}
These four theories may be further reduced  to $\SUSY =1, 2, 4$ and $8$ theories all with maximum rank $r=7$ in $D=4$, corresponding to compactification on $X^{(8-{\cal N})}\times T^{({\cal N}-1)}$.  Denote  the betti numbers of $X^7$, $X^6$, $X^4$ by $b$, $c$, $d$, respectively.  The betti numbers of $S^1$ are $(1,1)$,  of $T^3$ are $(1,3,3,1)$, of $T^4$ are $(1,4,6,4,1)$, of $T^7$ are $(1,7,21,35,21,7,1)$, so we have
\begin{equation}\label{bet}
\begin{split}
X^7&:(b_0 ,b_1, b_2, b_3)\\
X^6 \times S^1&: (c_0, c_0+c_1, c_1+c_2,c_2+c_3)\\
X^4 \times T^3&: (d_0, 3d_0+d_1, 3d_0+3d_1+d_2, d_0+4d_1+3d_2)\\
\end{split}
\end{equation}
The number of fields in $D=4$ is given by  Table \ref{D=4}. 

\begin{table}[h!]
$\begin{array}{llrrrrrrrrrr}
&Field &f&360A&X^7& X^6 \times S^1&X^4 \times T^3 &T^7 \\
&&&&&&&\\
\bigskip
&&&&\\
g_{MN}&g_{\mu\nu}&2&848&b_0&c_0&d_0&1\\
~&{\cal A}_{\mu}&2&-52&b_1&c_0+c_1&3d_0+d_1&7\\
~&{\cal A}&1&4&-b_1+b_3 &-c_0-c_1+c_2+c_3&-2d_0+3d_1+3d_2&28\\
\psi_{M}&\psi_{\mu}&2&-233&b_0+b_1&2c_0+c_1&4d_0+d_1&8\\
~&\chi&2&7&b_2+b_3&c_1+2c_2+c_3&4d_0+7d_1+4d_2&56\\
A_{MNP}&A_{\mu\nu\rho}&0&-720&b_0&d_0&c_0&1\\
~&A_{\mu\nu}&2&364&b_1&c_0+c_1&3d_0+d_1&7\\
~&A_{\mu}&2&-52&b_2&c_1+c_2&3d_0+3d_1+d_2&21\\
~&A&1&4&b_3&c_2+c_3&d_0+4d_1+3d_2&35\\
&&&&&\\
&&&&&&&\\
&&&&A=-\rho/24&A=-\chi/24&A=0&A=0&\\
\end{array}$
\caption{ $X^7, X^6\times S^1, X^4 \times T^3, T^7$ compactification of D=11 supergravity.}
\label{D=4}
\end{table}

\FloatBarrier
\subsection{Self-mirror with vanishing trace anomaly}
Finally, we note that in \cite{Duff:2010ss} we defined a generalized mirror symmetry
\begin{equation}
(b_0,  b_1, b_2, b_3) \rightarrow (b_0, b_1,  b_2 -\rho/2, b_3+\rho/2)
\label{eq:mirror}
\end{equation}
under which
\begin{equation}
\rho \equiv 7b_0-5b_1+3b_2 +b_3
\end{equation}
changes sign
\begin{equation}
\rho \rightarrow -\rho
\end{equation}
Moreover the quantity $\rho$ also shows up  in the on-shell trace anomaly (before dualization),  which is given by
\begin{equation}
g_{\mu\nu}<T^{\mu\nu}>=A \frac{1}{32\pi^2}R^{*}{}^{\mu\nu\rho\sigma}R^{*}{}_{\mu\nu\rho\sigma}
\end{equation}
The value of the $A$ coefficients for each field  is given in Table \ref{D=4} that shows compactification on $X^7$, $X^6 \times S^1$, $X^4 \times T^3$ and $T^7$. We adopt the interpretation of \cite{Duff:1980qv}  that assigns different anomalies to $A_{\mu\nu}$ and ${\cal A}$ even though they are naively dual to one another and nonzero anomaly to $A_{\mu\nu\rho}$. 
Remarkably, we find that the total anomaly depends on $\rho$
\begin{equation}
A=-\frac{1}{24}\rho
\end{equation}
So the anomaly flips sign under generalized mirror symmetry and vanishes for generalized self-mirror theories. 

In the case of  $({\cal N}=1,D=11)$ on $X^6 \times S^1$, or equivalently  (Type IIA, D=10) on $X^6$,
\begin{equation}
A=-\frac{1}{24}\chi
\end{equation}
where $\chi$ is the Euler number of $X^6$.

Here we group the individual fields into supermultiplets as shown in Tables \ref{Y7} to \ref{Y0}.
\begin{table}[h!]
$\begin{array}{llrrrrrrrrrr}
{\cal N}=1~&multiplet  &f&360A&{\cal N}=b_0+b_1 &  {\cal N}=8 &  {\cal N}=1 & \\

&&&&&&&\\
\bigskip
graviton&(g_{\mu\nu};\psi_{\mu}; A_{\mu\nu\rho})&2+2&-105&b_0&1&1\\
gravitino&({\cal A}_{\mu}; \psi_{\mu}) &2+2&-285&b_1&7&0 \\
&&&&&&&\\
vector&(\chi; A_{\mu})&2+2&-45&b_2&21&0\\
&&&&&&&\\
 WZ &({\cal A}; \chi; A)&2+2&15&-b_1+b_3&28&7\\
&&&&&&\\
linear&(\chi; A_{\mu\nu}, A)&2+2&375&b_1&7&0\\
&&&&&&\\
total ~f &&&&4(b_0+b_1+b_2+b_3)&256&32\\
&&&&&&\\
total ~A &&&&-(7b_0-5b_1+3b_2-b_3)/24&0&0\\
\end{array}$
\caption{ The $D=4$ multiplets in an  ${\cal N}$=1 basis. }
\label{Y7}
\end{table}

\begin{table}[h!]
$\begin{array}{llrrrrrrrrrrrrr}
{\cal N}=2~& multiplet  &f&360A&{\cal N}=2c_0+c_1 &  {\cal N}=8 &  {\cal N}=2 & \\
&&&&&&&\\
\bigskip
graviton&(g_{\mu\nu}, {\cal A}_{\mu}; 2\psi_{\mu}; A_{\mu\nu\rho})&4+4&-390&c_0&1&1\\
gravitino&({\cal A}_{\mu}; \psi_{\mu}, \chi; A_{\mu}) &4+4&-330&c_1&6&0 \\
&&&&&&&\\
vector&({\cal A}, 2\chi; A_{\mu}, A)&4+4&-30&c_2&15&3\\
&&&&&&&\\
hyper&(2{\cal A}; 2\chi; 2A)&4+4&30&-c_0-c_1+c_3/2&3&3\\
&&&&&&\\
linear&({\cal A}; 2\chi; A_{\mu\nu}, 2A)&4+4&390&c_0+c_1&7&1\\
&&&&&&\\
total ~f &&&&4(2c_0+2c_1+2c_2+c_3)&256&64\\
&&&&&&\\
total ~A &&&&-(2c_0-2c_1+2c_2-c_3)/24&0&0\\
\end{array}$
\caption{  The $D=4$ multiplets in an  ${\cal N}$=2 basis.}
\label{Y6}
\end{table}

\begin{table}[h!]
$\begin{array}{llrrrrrrrrrr}
{\cal N}=4~  & multiplet&f&360A& {\cal N}=4d_0+d_1 &  {\cal N}=8 &  {\cal N}=4 & \\
&&&&&&&\\
\bigskip
graviton&(g_{\mu\nu}, 3 {\cal A}_{\mu},  {\cal A}, 4 \psi_{\mu}, 4 \chi, A_{\mu\nu\rho}, 3  A_{\mu}, A)&16+16&-1080&d_0&1&1\\
gravitino&({\cal A}_{\mu},  3{\cal A},  \psi_{\mu}, 7 \chi,  A_{\mu\nu}, 3A_{\mu},4A) &16+16&0&d_1&4&0 \\
&&&&&&&\\
vector&(3{\cal A};  4 \chi; A_{\mu}, 3 A)&8+8&0&-3d_0+d_2&3&3\\
&&&&&\\
2-form &(2{\cal A};  4 \chi; A_{\mu\nu},A_{\mu}, 3 A)&8+8&360&3d_0&3&3\\
&&&&&&&\\
total ~f &&&&16(2d_0+2d_1+d_2)&256&128\\
&&&&&&\\
total ~A &&&&0&0&0\\
\end{array}$
\caption{  The $D=4$ multiplets in an  ${\cal N}$=4 basis.}
\label{Y4}
\end{table}

\begin{table}[h!]
$\begin{array}{llrrrrrrrrrr}
{\cal N}=8~  & multiplet&f&360A&{\cal N}=8  \\
&&&&&&&\\
\bigskip
graviton&(g_{\mu\nu}, 7{\cal A}_{\mu}, 28 {\cal A}; 8\psi_{\mu}, 56\chi; A_{\mu\nu\rho}, 7A_{\mu\nu},21 A_{\mu},35 A)
&256&0&1&\\
&&&&&&&\\  
total ~f &&&&256\\
&&&&&&&\\
total ~A &&&&0&\\
\end{array}$
\caption{  The $D=4$ multiplets in an  ${\cal N}$=8 basis.}
\label{Y0}
\end{table}

\FloatBarrier
\subsection{Cosets}
The $D=4$ scalar cosets, before and after dualization are given in Tables \ref{1cosetsD=4} and  \ref{2cosetsD=4}

\begin{table}[h!]
$\begin{array}{llllllllcccccc}
theory&charges&G/H&dim&rank\\
&&&&&\\
 N=8   & 32   &SO(1,1)\times SL(7,R)/SO(7) \ltimes R^{35}&63&7 \\
N=4&16&SL(2)/SO(2) \times SO(6,3)/[SO(6) \times SO(3)] \times SL(3,R)/SO(3)  \ltimes R^{9}&35&7\\
N=2&8&SO(1,1) \times SL(2)^3/SO(2)^3 \times SO(4,3)/[SO(4) \times SO(3)]\ltimes R^{2}&21&7\\
N=1&4&SL(2,R)^7/SO(2)^7 &14&7 \\
\end{array}$
\label{1cosetsD=4}
\caption{$D=4$ cosets before dualization}
\end{table}

\begin{table}[h!]
$\begin{array}{llllllllcccccc}
theory&charges&G/H&dim&rank\\
&&&&&\\
 N=8   & 32   &E_7/SU(8)&70&7 \\
N=4&16&SL(2)/SO(2) \times SO(6,6)/SO(6)^2&38&7\\
N=2&8&SL(2)^3/SO(2)^3 \times SO(4,4)/SO(4)^2&22&7\\
N=1&4&SL(2,R)^7/SO(2)^7&14 &7\\
\end{array}$
\label{2cosetsD=4}
\caption{$D$=4 cosets after dualization}
\end{table}

\FloatBarrier
\subsection{Fano plane and $\mathds{O, C, H, R}$}
\label{Fano}
 
\begin{figure}[h!]
\centering
\includegraphics[width=0.5\textwidth]{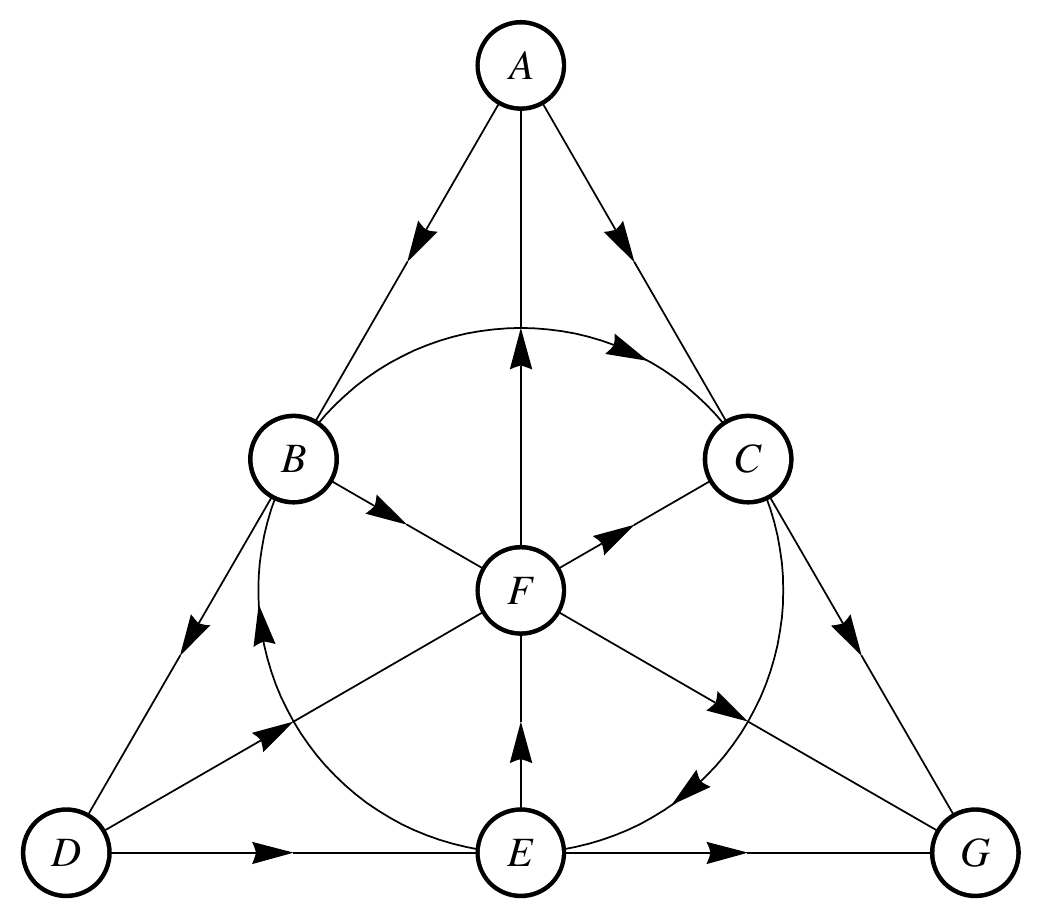}
\caption[Fano plane]{The Fano plane has seven points and seven lines (the circle
counts as a line) with three points on every line and three lines through every point. The points $A,B,C,D,E,F,G$ represent the seven qubits  and the seven lines $ABD, BCE, CDF, DEG, EFA, FGB, GAC$ represent  the tripartite entanglement.}
\label{fano}
\end{figure}

  Next we turn to the black-hole/qubit correspondence 
  \cite{Duff:2006uz,Kallosh:2006zs,Levay:2006kf,Duff:2006ue,Levay:2006pt,Borsten:2008wd}.  The number of electric and magnetic black hole charges of these ${\cal N}=8, 4, 2, 1$ theories are 56, 24, 8, 0, respectively.  These correspond to 7, 3, 1, 0 lines of the Fano plane of Fig \ref{fano}, which in turn admit an interpretation in terms of entangled qubits, as may be seen  by writing them all in an $SL(2)^7$ basis:

\begin{itemize}
\item{$\mathcal{N}=8$}
\begin{equation}
E_{7(7)} \supset SL(2)_A \times SL(2)_B \times SL(2)_C \times SL(2)_D \times SL(2)_E \times SL(2)_F \times SL(2)_G,
\end{equation}
and the \textbf{56} decomposes as
\begin{equation}\label{eq:56Decomp}
\begin{split}
\textbf{56} \to &\phantom{+\ }(\textbf{2},\textbf{2},\textbf{1},\textbf{2},\textbf{1},\textbf{1},\textbf{1}) \\
&+ (\textbf{1},\textbf{2},\textbf{2},\textbf{1},\textbf{2},\textbf{1},\textbf{1}) \\
&+ (\textbf{1},\textbf{1},\textbf{2},\textbf{2},\textbf{1},\textbf{2},\textbf{1}) \\
&+ (\textbf{1},\textbf{1},\textbf{1},\textbf{2},\textbf{2},\textbf{1},\textbf{2}) \\
&+ (\textbf{2},\textbf{1},\textbf{1},\textbf{1},\textbf{2},\textbf{2},\textbf{1}) \\
&+ (\textbf{1},\textbf{2},\textbf{1},\textbf{1},\textbf{1},\textbf{2},\textbf{2}) \\
&+ (\textbf{2},\textbf{1},\textbf{2},\textbf{1},\textbf{1},\textbf{1},\textbf{2}).
\end{split}
\end{equation}
corresponding to the seven lines of the Fano plane describing a tripartite entanglement of seven qubits (Alice, Bob, Charlie, Daisy, Emma, Fred and George):

\begin{equation}\label{56}
\begin{split}
\textbf{$|\psi\rangle_{56}$} = &\phantom{+\ }a_{ABD}|ABD\rangle \\
&+ b_{BCE}|BCE\rangle \\
&+ c_{CDF}|CDF\rangle  \\
&+ d_{DEG}|DEG\rangle \\ 
&+ e_{EFA}|EFA\rangle \\ 
&+ f_{FGB}|FGB\rangle \\
&+ g_{GAC}|GAC\rangle
\end{split}
\end{equation}

\item{$\mathcal{N}=4$}
\begin{equation}
SL(2)_A \times SO(6,6) \supset SL(2)_A \times SL(2)_B \times SL(2)_C \times SL(2)_D \times SL(2)_E \times SL(2)_F \times SL(2)_G,
\end{equation}
and the (\textbf{2,12}) decomposes as
\begin{equation}\label{eq:56Decomp}
\begin{split}
 (\textbf{2,12})  \to &\phantom{+\ }
 (\textbf{2},\textbf{2},\textbf{1},\textbf{2},\textbf{1},\textbf{1},\textbf{1}) \\
&+ (\textbf{2},\textbf{1},\textbf{1},\textbf{1},\textbf{2},\textbf{2},\textbf{1}) \\
&+ (\textbf{2},\textbf{1},\textbf{2},\textbf{1},\textbf{1},\textbf{1},\textbf{2}).
\end{split}
\end{equation}
corresponding to the three lines of the Fano plane describing a tripartite entanglement of Alice with Bob and Daisy, Alice with Emma and Fred, Alice with Charlie and George:
\begin{equation}\label{24}
\begin{split}
\textbf{$|\psi\rangle_{24}$} = &\phantom{+\ }a_{ABD}|ABD\rangle \\
&+ e_{EFA}|EFA\rangle \\ 
&+ g_{GAC}|GAC\rangle
\end{split}
\end{equation}

\item{$\mathcal{N}=2$}
\begin{equation}
SL(2)_A \times SL(2)_B \times SL(2)_D \times SO(4,4)\supset SL(2)_A \times SL(2)_B \times SL(2)_C \times SL(2)_D \times SL(2)_E \times SL(2)_F \times SL(2)_G,
\end{equation}
and the $(\textbf{2,2,2,1})$ decomposes as
\begin{equation}\label{eq:56Decomp}
\begin{split}
(\textbf{2,2,2,1}) \to &\phantom{+\ }(\textbf{2},\textbf{2},\textbf{1},\textbf{2},\textbf{1},\textbf{1},\textbf{1}) \\
\end{split}
\end{equation}
corresponding to the one line of the Fano plane describing a tripartite entanglement of three qubits, Alice, Bob and Daisy:
\begin{equation}
    |\psi\rangle_{8} = \begin{array}{r@{\ }c@{}l}
                       & a_{ABD} & |ABD\rangle \\
                     
                   \end{array}
\end{equation}

\item{$\mathcal{N}=1$}
\[
SL(2)_A \times SL(2)_B \times SL(2)_C \times SL(2)_D \times SL(2)_E \times SL(2)_F \times SL(2)_G
\]
\begin{equation}
\supset SL(2)_A \times SL(2)_B \times SL(2)_C \times SL(2)_D \times SL(2)_E \times SL(2)_F \times SL(2)_G,
\end{equation}

\begin{equation}
\textbf{(0,0,0,0,0,0,0)} \to  \textbf{(0,0,0,0,0,0,0)} 
\end{equation}
corresponding to no lines.

\end{itemize}
The black hole entropies are given by the qubit entanglement measures which are quartic polynomials in the $a,b,c,d,e,f,g$ coefficients, namely Cartan's $E_7$ invariant, the analogous $SL(2) \times SO(6,6)$ invariant and Cayley's $SL(2)^3$ hyperdeterminant.

Since the $7, 3, 1, 0$ lines of the Fano plane also describe the multiplication table of the octonions, quaternions, complex and real, it was conjectured in \cite{Borsten:2008wd} that there is an $\mathds{O, H, C, R}$ interpretation not just for the charges but for the entire theories. Their field content and trace anomalies are given in Tables \ref{curious1} and \ref{curious2}.

\begin{table}[h!]
$\begin{array}{lrrrrrrrrrr}
Field & f&360A &\mathds{A}&\mathds{O}  & \mathds{H}  & \mathds{C}  &\mathds{R}\\
\bigskip
&&&&&\\
g_{\mu\nu}&2&848&1&1&1&1&1\\
{\cal A}_{\mu}&2&-52&{\cal N}-1&7&3&1&0\\
\vec\phi&1&4&7&7&7&7&7\\
{\cal A}&1&4&3({\cal N}-1)&21&9&3&0\\
\psi_{\mu}&2&-233&{\cal N}&8&4&2&1\\
\chi&2&7&7{\cal N}&56&28&14&7\\
A_{\mu\nu\rho}&0&-720&1&1&1&1&1\\
A_{\mu\nu}&1&364&{\cal N}-1&7&3&1&0\\
A_{\mu}&2&-52&3({\cal N}-1)&21&9&3&0\\
A&1&4&4{\cal N}+3&35&19&11&7\\
&&&&&&\\
total ~f &&&32{\cal N}&256&128&64&32\\
&&&&&&&&\\
total~A&&&0&0&0&0&0\\
\end{array}$
\caption{Vanishing anomaly in $\mathds{O}$, $\mathds{H}$, $\mathds{C}$ $\mathds{R}$ theories.}
\label{curious1}
\end{table}

\begin{table}[h!]
$\begin{array}{llrrrrrrrrrr}
{\cal N}=1&multiplet& f&360A&\mathds{A} & \mathds{O}& \mathds{H}  & \mathds{C}  &\mathds{R}\\

&&&&&&&\\
\bigskip
graviton&(g_{\mu\nu};\psi_{\mu}; A_{\mu\nu\rho})&2+2&-105&1&1&1&1&1\\
gravitino&({\cal A}_{\mu}; \psi_{\mu}) &2+2&-285&{\cal N}-1&7&3&1&0 \\
&&&&&&&\\
vector&(\chi; A_{\mu})&2+2&-45&3({\cal N}-1)&21&9&3&0\\
&&&&&&&\\
 WZ_{\phi} &(\vec\phi; \chi; A)&2+2&15&7&7&7&7&7\\
 &&&&&&&\\
 WZ_{{\cal A}} &({\cal A}; \chi; A)&2+2&15&3({\cal N}-1)&21&9&3&0\\
&&&&&&\\
linear&(\chi; A_{\mu\nu}, A)&2+2&375&{\cal N}-1&7&3&1&0\\
&&&&&&\\
total ~f &&&&32{\cal N}&256&128&64&32\\
&&&&&&\\
total ~A &&&&0&0&0&0&0\\
\end{array}$
\caption{ The $D=4$ multiplets in an  ${\cal N}$=1 basis from $X^7$ with $(b_0,b_1,b_2,b_3)=(1,{\cal N}-1,3{\cal N}-3, 4{\cal N}+3)$}
\label{curious2}
\end{table}

\FloatBarrier
\section{$\mathcal{N}=2,4,8,16$ in $D=3$}
\subsection{Compactifications}

Consider Type IIA in $D=10$.  In the NS sector we have the fields
$(g_{MN}, \Phi; \psi_M, \chi; {A}_{MN})$ with $f=64+64$; in the R-R we have the fields $({\cal A}_M;\psi_M, \chi; A_{MNP})$ also with $f=64+64$.  We compactify on generic $X^7$ with independent betti numbers $(b_0,b_1,b_2,b_3)$, $X^6 \times S^1$ with independent $X^6$ betti numbers $(c_0,c_1,c_2,c_3)$,  $X^4 \times S^3$ with independent $X^4$ betti numbers $(d_0,d_1,d_2)$ and on $T^7$ with $(1, 7, 21, 35)$. The results for NS and RR combined are shown in Table \ref{D=4A}. In Table \ref{Y8}, we group into ${\cal N}=2$ multiplets.

\begin{table}[h!]
$\begin{array}{llrrrrrrrrrr}
&Field &f&X^7& X^6 \times S^1&X^4 \times T^3 &T^7 \\
&&&&&&&\\
\bigskip
&&&&\\
g_{MN}&g_{\mu\nu}&0&b_0&c_0&d_0&1\\
~&{\cal A}_{\mu}&1&b_0+b_1&2c_0+c_1&4d_0+d_1&8\\
~&{\cal A}&1&b_3 &c_2+c_3&d_0+4d_1+3d_2&35\\
\Phi&\Phi&1&b_0&c_0&d_0&1\\
\psi_{M}&\psi_{\mu}&0&2b_0+2b_1&4c_0+2c_1&8d_0+2d_1&16\\
~&\chi&1&2b_0+2b_1+2b_2+2b_3&4c_0+4c_1+4c_2+2c_3&16d_0+16d_1+8d_2&128\\
A_{MNP}&A_{\mu\nu\rho}&0&b_0&c_0&d_0&1\\
~&A_{\mu\nu}&0&b_0+b_1&2c_0+c_1&3d_0+d_1&8\\
~&A_{\mu}&1&b_1+b_2&c_0+2c_1+c_2&6d_0+4d_1+d_2&28\\
~&A&1&b_2+b_3&c_1+2c_2+c_3&4d_0+7d_1+4d_2&56\\
&&&&&\\
&&&&&&&\\
total~f&&&4(b_0+b_1+b_2+b_3)&4(2c_0+2c_1+2c_2+c_3)&16(2d_0+2d_1+d_2)&256&\\
&&&&&&&
\end{array}$
\caption{ $X^7, X^6\times S^1, X^4 \times T^3, T^7$ compactification of Type IIA}
\label{D=4A}
\end{table}

\begin{table}[h!]
$\begin{array}{llrrrrrrrrrr}
{\cal N}=2~multiplet & content &f&{\cal N}=2b_0+2b_1&  {\cal N}=16&{\cal N}=8&{\cal N}=4&  {\cal N}=2 & \\
&&&&&&&\\
\bigskip
graviton&(g_{\mu\nu}, {\cal A}_{\mu}, {\cal A}; 2\psi_{\mu}, 2\chi;  A_{\mu\nu\rho}, A_{\mu\nu})&2+2&b_0&1&1&1&1\\
gravitino&({\cal A}_{\mu},{\cal A}; 2\psi_{\mu}, 2\chi)&2+2&b_1&7&3&1&0\\
&&&&&&&\\
vector&(2\chi; A_{\mu}, A)&2+2&b_2&21&9&3&0\\
&&&&&&&\\
 hyper &({\cal A}; 2\chi; A)&2+2&-b_1+b_3&28&16&10&7\\
&&&&&&\\
linear&(2\chi; A_{\mu\nu}, A_{\mu}, A)&2+2&b_1&7&3&1&0\\
&&&&&&\\
total~f&&&4(b_0+b_1+b_2+b_3)&256&128&64&32\\
\end{array}$
\caption{ The $D=3$ multiplets in an ${\mathcal N}=2$ basis}
\label{Y8}
\end{table}

\FloatBarrier
\subsection{Cosets}
The $D=3$  scalar cosets after dualization for spacelike and timelike reductions are given in Tables \ref{2cosetsD=3} and \ref{2cosetsD=3'}
\begin{table}[h!]
$\begin{array}{lllrrllcccccc}
theory&&G/H&dim&rank\\
&&&&&\\
{\cal N}=16   &    &E_8/SO(16)&128&8 \\
{\cal N}=8&&SO(8,8)/SO(8)^2&64&8\\
{\cal N}=4&&SO(4,4)^2/SO(4)^4&32&8\\
{\cal N}=2&&SL(2,R)^8/SO(2)^8&16&8 \\
\end{array}$
\label{2cosetsD=3}
\caption{$D=3$ cosets after dualization}
\end{table}

\begin{table}[h!]
$\begin{array}{lllrrllcccccc}
theory&&G/H&dim&rank\\
&&&&&\\
{\cal N}=16   &    &E_8/SO^*(16)&128&8 \\
{\cal N}=8&&SO(8,8)/SO(4,4)^2&64&8\\
{\cal N}=4&&SO(4,4)^2/SO(2,2)^4&32&8\\
{\cal N}=2&&SL(2,R)^8/SO(1,1)^8&16&8 \\
\end{array}$
\label{2cosetsD=3'}
\caption{$D=3$ cosets from timelike reduction}
\end{table}

\section{Acknowledgments}

Conversations and correspondence with Leron Borsten, Duminda Dahanayke, Phil Gibbs, Alessio Marrani, Chris Pope, William Rubens,  Ashoke Sen, Ergin Sezgin, Samson Shatashvili, John Schwarz and Edward Witten are much appreciated. MJD is supported in part by the STFC under rolling Grant No. ST/G000743/1.  S. F is supported by the ERC Advanced Grant no 226455,``Supersymmetry, Quantum Gravity and Gauge Fields" and in part by DOE Grant DE-FG03-91ER40662.  MJD is grateful for hospitality at the CERN theory division, where he was supported by the above ERC Advanced Grant, and at the Mitchell Institute for Fundamental Physics and the Institute for Quantum Studies, Texas A\&M University.

\FloatBarrier
\bibliographystyle{apsrev}
\bibliography{arxiv,notarxiv}

\begin{thebibliography}{41}
\expandafter\ifx\csname natexlab\endcsname\relax\def\natexlab#1{#1}\fi
\expandafter\ifx\csname bibnamefont\endcsname\relax
  \def\bibnamefont#1{#1}\fi
\expandafter\ifx\csname bibfnamefont\endcsname\relax
  \def\bibfnamefont#1{#1}\fi
\expandafter\ifx\csname citenamefont\endcsname\relax
  \def\citenamefont#1{#1}\fi
\expandafter\ifx\csname url\endcsname\relax
  \def\url#1{\texttt{#1}}\fi
\expandafter\ifx\csname urlprefix\endcsname\relax\def\urlprefix{URL }\fi
\providecommand{\bibinfo}[2]{#2}
\providecommand{\eprint}[2][]{\url{#2}}

\bibitem[{\citenamefont{Green and Schwarz}(1984)}]{Green:1983wt}
\bibinfo{author}{\bibfnamefont{M.~B.} \bibnamefont{Green}} \bibnamefont{and}
  \bibinfo{author}{\bibfnamefont{J.~H.} \bibnamefont{Schwarz}},
  \bibinfo{journal}{Phys. Lett.} \textbf{\bibinfo{volume}{B136}},
  \bibinfo{pages}{367} (\bibinfo{year}{1984}).

\bibitem[{\citenamefont{Achucarro et~al.}(1987)\citenamefont{Achucarro, Evans,
  Townsend, and Wiltshire}}]{Achucarro:1987nc}
\bibinfo{author}{\bibfnamefont{A.}~\bibnamefont{Achucarro}},
  \bibinfo{author}{\bibfnamefont{J.~M.} \bibnamefont{Evans}},
  \bibinfo{author}{\bibfnamefont{P.~K.} \bibnamefont{Townsend}},
  \bibnamefont{and} \bibinfo{author}{\bibfnamefont{D.~L.}
  \bibnamefont{Wiltshire}}, \bibinfo{journal}{Phys. Lett.}
  \textbf{\bibinfo{volume}{B198}}, \bibinfo{pages}{441} (\bibinfo{year}{1987}).

\bibitem[{\citenamefont{Duff and Lu}(1993)}]{Duff:1992hu}
\bibinfo{author}{\bibfnamefont{M.~J.} \bibnamefont{Duff}} \bibnamefont{and}
  \bibinfo{author}{\bibfnamefont{J.~X.} \bibnamefont{Lu}},
  \bibinfo{journal}{Nucl. Phys.} \textbf{\bibinfo{volume}{B390}},
  \bibinfo{pages}{276} (\bibinfo{year}{1993}), \eprint{hep-th/9207060}.

\bibitem[{\citenamefont{Duff et~al.}(1995)\citenamefont{Duff, Khuri, and
  Lu}}]{Duff:1994an}
\bibinfo{author}{\bibfnamefont{M.~J.} \bibnamefont{Duff}},
  \bibinfo{author}{\bibfnamefont{R.~R.} \bibnamefont{Khuri}}, \bibnamefont{and}
  \bibinfo{author}{\bibfnamefont{J.~X.} \bibnamefont{Lu}},
  \bibinfo{journal}{Phys. Rept.} \textbf{\bibinfo{volume}{259}},
  \bibinfo{pages}{213} (\bibinfo{year}{1995}), \eprint{hep-th/9412184}.

\bibitem[{\citenamefont{Duff et~al.}(1987)\citenamefont{Duff, Howe, Inami, and
  Stelle}}]{Duff:1987bx}
\bibinfo{author}{\bibfnamefont{M.~J.} \bibnamefont{Duff}},
  \bibinfo{author}{\bibfnamefont{P.~S.} \bibnamefont{Howe}},
  \bibinfo{author}{\bibfnamefont{T.}~\bibnamefont{Inami}}, \bibnamefont{and}
  \bibinfo{author}{\bibfnamefont{K.~S.} \bibnamefont{Stelle}},
  \bibinfo{journal}{Phys. Lett.} \textbf{\bibinfo{volume}{B191}},
  \bibinfo{pages}{70} (\bibinfo{year}{1987}).

\bibitem[{\citenamefont{Bergshoeff et~al.}(1987)\citenamefont{Bergshoeff,
  Sezgin, and Townsend}}]{Bergshoeff:1987cm}
\bibinfo{author}{\bibfnamefont{E.}~\bibnamefont{Bergshoeff}},
  \bibinfo{author}{\bibfnamefont{E.}~\bibnamefont{Sezgin}}, \bibnamefont{and}
  \bibinfo{author}{\bibfnamefont{P.~K.} \bibnamefont{Townsend}},
  \bibinfo{journal}{Phys. Lett.} \textbf{\bibinfo{volume}{B189}},
  \bibinfo{pages}{75} (\bibinfo{year}{1987}).

\bibitem[{\citenamefont{Duff and Stelle}(1991)}]{Duff:1990xz}
\bibinfo{author}{\bibfnamefont{M.~J.} \bibnamefont{Duff}} \bibnamefont{and}
  \bibinfo{author}{\bibfnamefont{K.~S.} \bibnamefont{Stelle}},
  \bibinfo{journal}{Phys. Lett.} \textbf{\bibinfo{volume}{B253}},
  \bibinfo{pages}{113} (\bibinfo{year}{1991}).

\bibitem[{\citenamefont{Bergshoeff et~al.}(1986)\citenamefont{Bergshoeff,
  Sezgin, and Townsend}}]{Bergshoeff:1985su}
\bibinfo{author}{\bibfnamefont{E.}~\bibnamefont{Bergshoeff}},
  \bibinfo{author}{\bibfnamefont{E.}~\bibnamefont{Sezgin}}, \bibnamefont{and}
  \bibinfo{author}{\bibfnamefont{P.~K.} \bibnamefont{Townsend}},
  \bibinfo{journal}{Phys. Lett.} \textbf{\bibinfo{volume}{B169}},
  \bibinfo{pages}{191} (\bibinfo{year}{1986}).

\bibitem[{\citenamefont{Ovrut and Waldram}(1997)}]{Ovrut:1997ur}
\bibinfo{author}{\bibfnamefont{B.~A.} \bibnamefont{Ovrut}} \bibnamefont{and}
  \bibinfo{author}{\bibfnamefont{D.}~\bibnamefont{Waldram}},
  \bibinfo{journal}{Nucl. Phys.} \textbf{\bibinfo{volume}{B506}},
  \bibinfo{pages}{236} (\bibinfo{year}{1997}), \eprint{hep-th/9704045}.

\bibitem[{\citenamefont{Baez and Huerta}(2009)}]{Baez:2009xt}
\bibinfo{author}{\bibfnamefont{J.~C.} \bibnamefont{Baez}} \bibnamefont{and}
  \bibinfo{author}{\bibfnamefont{J.}~\bibnamefont{Huerta}}
  (\bibinfo{year}{2009}), \eprint{0909.0551}.

\bibitem[{\citenamefont{Baez and Huerta}(2010)}]{Baez:2010ye}
\bibinfo{author}{\bibfnamefont{J.~C.} \bibnamefont{Baez}} \bibnamefont{and}
  \bibinfo{author}{\bibfnamefont{J.}~\bibnamefont{Huerta}}
  (\bibinfo{year}{2010}), \eprint{1003.3436}.

\bibitem[{\citenamefont{Cremmer et~al.}(1978)\citenamefont{Cremmer, Julia, and
  Scherk}}]{Cremmer:1978km}
\bibinfo{author}{\bibfnamefont{E.}~\bibnamefont{Cremmer}},
  \bibinfo{author}{\bibfnamefont{B.}~\bibnamefont{Julia}}, \bibnamefont{and}
  \bibinfo{author}{\bibfnamefont{J.}~\bibnamefont{Scherk}},
  \bibinfo{journal}{Phys. Lett.} \textbf{\bibinfo{volume}{B76}},
  \bibinfo{pages}{409} (\bibinfo{year}{1978}).

\bibitem[{\citenamefont{Ferrara and Kounnas}(1989)}]{Ferrara:1989nm}
\bibinfo{author}{\bibfnamefont{S.}~\bibnamefont{Ferrara}} \bibnamefont{and}
  \bibinfo{author}{\bibfnamefont{C.}~\bibnamefont{Kounnas}},
  \bibinfo{journal}{Nucl. Phys.} \textbf{\bibinfo{volume}{B328}},
  \bibinfo{pages}{406} (\bibinfo{year}{1989}).

\bibitem[{\citenamefont{Sen and Vafa}(1995)}]{Sen:1995ff}
\bibinfo{author}{\bibfnamefont{A.}~\bibnamefont{Sen}} \bibnamefont{and}
  \bibinfo{author}{\bibfnamefont{C.}~\bibnamefont{Vafa}},
  \bibinfo{journal}{Nucl. Phys.} \textbf{\bibinfo{volume}{B455}},
  \bibinfo{pages}{165} (\bibinfo{year}{1995}), \eprint{hep-th/9508064}.

\bibitem[{\citenamefont{Joyce}(1996{\natexlab{a}})}]{Joyce:1996a}
\bibinfo{author}{\bibfnamefont{D.}~\bibnamefont{Joyce}}, \bibinfo{journal}{J.
  Differential Geometry} \textbf{\bibinfo{volume}{43}}
  (\bibinfo{year}{1996}{\natexlab{a}}).

\bibitem[{\citenamefont{Joyce}(1996{\natexlab{b}})}]{Joyce:1996b}
\bibinfo{author}{\bibfnamefont{D.}~\bibnamefont{Joyce}}, \bibinfo{journal}{J.
  Differential Geometry} \textbf{\bibinfo{volume}{43}}, \bibinfo{pages}{329}
  (\bibinfo{year}{1996}{\natexlab{b}}).

\bibitem[{\citenamefont{Gaberdiel and Kaste}(2004)}]{Gaberdiel:2004vx}
\bibinfo{author}{\bibfnamefont{M.~R.} \bibnamefont{Gaberdiel}}
  \bibnamefont{and} \bibinfo{author}{\bibfnamefont{P.}~\bibnamefont{Kaste}},
  \bibinfo{journal}{JHEP} \textbf{\bibinfo{volume}{08}}, \bibinfo{pages}{001}
  (\bibinfo{year}{2004}), \eprint{hep-th/0401125}.

\bibitem[{\citenamefont{D'Auria et~al.}(2004)\citenamefont{D'Auria, Sommovigo,
  and Vaula}}]{D'Auria:2004yi}
\bibinfo{author}{\bibfnamefont{R.}~\bibnamefont{D'Auria}},
  \bibinfo{author}{\bibfnamefont{L.}~\bibnamefont{Sommovigo}},
  \bibnamefont{and} \bibinfo{author}{\bibfnamefont{S.}~\bibnamefont{Vaula}},
  \bibinfo{journal}{JHEP} \textbf{\bibinfo{volume}{11}}, \bibinfo{pages}{028}
  (\bibinfo{year}{2004}), \eprint{hep-th/0409097}.

\bibitem[{\citenamefont{Duff}(2007)}]{Duff:2006uz}
\bibinfo{author}{\bibfnamefont{M.~J.} \bibnamefont{Duff}},
  \bibinfo{journal}{Phys. Rev.} \textbf{\bibinfo{volume}{D76}},
  \bibinfo{pages}{025017} (\bibinfo{year}{2007}), \eprint{hep-th/0601134}.

\bibitem[{\citenamefont{Kallosh and Linde}(2006)}]{Kallosh:2006zs}
\bibinfo{author}{\bibfnamefont{R.}~\bibnamefont{Kallosh}} \bibnamefont{and}
  \bibinfo{author}{\bibfnamefont{A.}~\bibnamefont{Linde}},
  \bibinfo{journal}{Phys. Rev.} \textbf{\bibinfo{volume}{D73}},
  \bibinfo{pages}{104033} (\bibinfo{year}{2006}), \eprint{hep-th/0602061}.

\bibitem[{\citenamefont{L\'evay}(2006)}]{Levay:2006kf}
\bibinfo{author}{\bibfnamefont{P.}~\bibnamefont{L\'evay}},
  \bibinfo{journal}{Phys. Rev.} \textbf{\bibinfo{volume}{D74}},
  \bibinfo{pages}{024030} (\bibinfo{year}{2006}), \eprint{hep-th/0603136}.

\bibitem[{\citenamefont{Duff and Ferrara}(2007)}]{Duff:2006ue}
\bibinfo{author}{\bibfnamefont{M.~J.} \bibnamefont{Duff}} \bibnamefont{and}
  \bibinfo{author}{\bibfnamefont{S.}~\bibnamefont{Ferrara}},
  \bibinfo{journal}{Phys. Rev.} \textbf{\bibinfo{volume}{D76}},
  \bibinfo{pages}{025018} (\bibinfo{year}{2007}), \eprint{quant-ph/0609227}.

\bibitem[{\citenamefont{L\'evay}(2007)}]{Levay:2006pt}
\bibinfo{author}{\bibfnamefont{P.}~\bibnamefont{L\'evay}},
  \bibinfo{journal}{Phys. Rev.} \textbf{\bibinfo{volume}{D75}},
  \bibinfo{pages}{024024} (\bibinfo{year}{2007}), \eprint{hep-th/0610314}.

\bibitem[{\citenamefont{Borsten et~al.}(2009)\citenamefont{Borsten, Dahanayake,
  Duff, Ebrahim, and Rubens}}]{Borsten:2008wd}
\bibinfo{author}{\bibfnamefont{L.}~\bibnamefont{Borsten}},
  \bibinfo{author}{\bibfnamefont{D.}~\bibnamefont{Dahanayake}},
  \bibinfo{author}{\bibfnamefont{M.~J.} \bibnamefont{Duff}},
  \bibinfo{author}{\bibfnamefont{H.}~\bibnamefont{Ebrahim}}, \bibnamefont{and}
  \bibinfo{author}{\bibfnamefont{W.}~\bibnamefont{Rubens}},
  \bibinfo{journal}{Phys. Rep.} \textbf{\bibinfo{volume}{471}},
  \bibinfo{pages}{113} (\bibinfo{year}{2009}), \eprint{0809.4685}.

\bibitem[{\citenamefont{Kugo and Townsend}(1983)}]{Kugo:1982bn}
\bibinfo{author}{\bibfnamefont{T.}~\bibnamefont{Kugo}} \bibnamefont{and}
  \bibinfo{author}{\bibfnamefont{P.~K.} \bibnamefont{Townsend}},
  \bibinfo{journal}{Nucl. Phys.} \textbf{\bibinfo{volume}{B221}},
  \bibinfo{pages}{357} (\bibinfo{year}{1983}).

\bibitem[{\citenamefont{Evans}(1988)}]{Evans:1987tm}
\bibinfo{author}{\bibfnamefont{J.~M.} \bibnamefont{Evans}},
  \bibinfo{journal}{Nucl. Phys.} \textbf{\bibinfo{volume}{B298}},
  \bibinfo{pages}{92} (\bibinfo{year}{1988}).

\bibitem[{\citenamefont{Duff}(1988)}]{Duff:1987qa}
\bibinfo{author}{\bibfnamefont{M.~J.} \bibnamefont{Duff}},
  \bibinfo{journal}{Class. Quant. Grav.} \textbf{\bibinfo{volume}{5}},
  \bibinfo{pages}{189} (\bibinfo{year}{1988}).

\bibitem[{\citenamefont{Duff and Ferrara}(2010)}]{Duff:2010ss}
\bibinfo{author}{\bibfnamefont{M.~J.} \bibnamefont{Duff}} \bibnamefont{and}
  \bibinfo{author}{\bibfnamefont{S.}~\bibnamefont{Ferrara}}
  (\bibinfo{year}{2010}), \eprint{1009.4439}.

\bibitem[{\citenamefont{Shatashvili and Vafa}(1995)}]{Shatashvili:1994zw}
\bibinfo{author}{\bibfnamefont{S.~L.} \bibnamefont{Shatashvili}}
  \bibnamefont{and} \bibinfo{author}{\bibfnamefont{C.}~\bibnamefont{Vafa}},
  \bibinfo{journal}{Selecta Math.} \textbf{\bibinfo{volume}{1}},
  \bibinfo{pages}{347} (\bibinfo{year}{1995}), \eprint{hep-th/9407025}.

\bibitem[{\citenamefont{Acharya}(1998)}]{Acharya:1997rh}
\bibinfo{author}{\bibfnamefont{B.~S.} \bibnamefont{Acharya}},
  \bibinfo{journal}{Nucl. Phys.} \textbf{\bibinfo{volume}{B524}},
  \bibinfo{pages}{269} (\bibinfo{year}{1998}), \eprint{hep-th/9707186}.

\bibitem[{\citenamefont{Duff}(1977)}]{Duff:1977ay}
\bibinfo{author}{\bibfnamefont{M.~J.} \bibnamefont{Duff}},
  \bibinfo{journal}{Nucl. Phys.} \textbf{\bibinfo{volume}{B125}},
  \bibinfo{pages}{334} (\bibinfo{year}{1977}).

\bibitem[{\citenamefont{Duff}(1994)}]{Duff:1993wm}
\bibinfo{author}{\bibfnamefont{M.~J.} \bibnamefont{Duff}},
  \bibinfo{journal}{Class. Quant. Grav.} \textbf{\bibinfo{volume}{11}},
  \bibinfo{pages}{1387} (\bibinfo{year}{1994}), \eprint{hep-th/9308075}.

\bibitem[{\citenamefont{Duff and van Nieuwenhuizen}(1980)}]{Duff:1980qv}
\bibinfo{author}{\bibfnamefont{M.~J.} \bibnamefont{Duff}} \bibnamefont{and}
  \bibinfo{author}{\bibfnamefont{P.}~\bibnamefont{van Nieuwenhuizen}},
  \bibinfo{journal}{Phys. Lett.} \textbf{\bibinfo{volume}{B94}},
  \bibinfo{pages}{179} (\bibinfo{year}{1980}).

\bibitem[{\citenamefont{Levay}(2010)}]{Levay:2010ua}
\bibinfo{author}{\bibfnamefont{P.}~\bibnamefont{Levay}} (\bibinfo{year}{2010}),
  \eprint{1004.3639}.

\bibitem[{\citenamefont{Borsten
  et~al.}(2010{\natexlab{a}})\citenamefont{Borsten, Dahanayake, Duff, Marrani,
  and Rubens}}]{Borsten:2010db}
\bibinfo{author}{\bibfnamefont{L.}~\bibnamefont{Borsten}},
  \bibinfo{author}{\bibfnamefont{D.}~\bibnamefont{Dahanayake}},
  \bibinfo{author}{\bibfnamefont{M.~J.} \bibnamefont{Duff}},
  \bibinfo{author}{\bibfnamefont{A.}~\bibnamefont{Marrani}}, \bibnamefont{and}
  \bibinfo{author}{\bibfnamefont{W.}~\bibnamefont{Rubens}}
  (\bibinfo{year}{2010}{\natexlab{a}}), \eprint{1005.4915}.

\bibitem[{\citenamefont{Borsten et~al.}(2010{\natexlab{b}})}]{Borsten:2010ab}
\bibinfo{author}{\bibfnamefont{L.}~\bibnamefont{Borsten}} \bibnamefont{et~al.}
  (\bibinfo{year}{2010}{\natexlab{b}}).

\bibitem[{\citenamefont{Gibbs}(2010)}]{Gibbs:2010aa}
\bibinfo{author}{\bibfnamefont{P.}~\bibnamefont{Gibbs}} (\bibinfo{year}{2010}),
  \eprint{1009.0076}.

\bibitem[{\citenamefont{Cremmer
  et~al.}(1998{\natexlab{a}})\citenamefont{Cremmer, Julia, Lu, and
  Pope}}]{Cremmer:1997ct}
\bibinfo{author}{\bibfnamefont{E.}~\bibnamefont{Cremmer}},
  \bibinfo{author}{\bibfnamefont{B.}~\bibnamefont{Julia}},
  \bibinfo{author}{\bibfnamefont{H.}~\bibnamefont{Lu}}, \bibnamefont{and}
  \bibinfo{author}{\bibfnamefont{C.~N.} \bibnamefont{Pope}},
  \bibinfo{journal}{Nucl. Phys.} \textbf{\bibinfo{volume}{B523}},
  \bibinfo{pages}{73} (\bibinfo{year}{1998}{\natexlab{a}}),
  \eprint{hep-th/9710119}.

\bibitem[{\citenamefont{Cremmer
  et~al.}(1998{\natexlab{b}})\citenamefont{Cremmer, Julia, Lu, and
  Pope}}]{Cremmer:1998px}
\bibinfo{author}{\bibfnamefont{E.}~\bibnamefont{Cremmer}},
  \bibinfo{author}{\bibfnamefont{B.}~\bibnamefont{Julia}},
  \bibinfo{author}{\bibfnamefont{H.}~\bibnamefont{Lu}}, \bibnamefont{and}
  \bibinfo{author}{\bibfnamefont{C.~N.} \bibnamefont{Pope}},
  \bibinfo{journal}{Nucl. Phys.} \textbf{\bibinfo{volume}{B535}},
  \bibinfo{pages}{242} (\bibinfo{year}{1998}{\natexlab{b}}),
  \eprint{hep-th/9806106}.

\bibitem[{\citenamefont{Lu and Pope}(1996)}]{Lu:1995yn}
\bibinfo{author}{\bibfnamefont{H.}~\bibnamefont{Lu}} \bibnamefont{and}
  \bibinfo{author}{\bibfnamefont{C.~N.} \bibnamefont{Pope}},
  \bibinfo{journal}{Nucl. Phys.} \textbf{\bibinfo{volume}{B465}},
  \bibinfo{pages}{127} (\bibinfo{year}{1996}), \eprint{hep-th/9512012}.

\bibitem[{\citenamefont{Lavrinenko et~al.}(1997)\citenamefont{Lavrinenko, Lu,
  and Pope}}]{Lavrinenko:1996mp}
\bibinfo{author}{\bibfnamefont{I.~V.} \bibnamefont{Lavrinenko}},
  \bibinfo{author}{\bibfnamefont{H.}~\bibnamefont{Lu}}, \bibnamefont{and}
  \bibinfo{author}{\bibfnamefont{C.~N.} \bibnamefont{Pope}},
  \bibinfo{journal}{Nucl. Phys.} \textbf{\bibinfo{volume}{B492}},
  \bibinfo{pages}{278} (\bibinfo{year}{1997}), \eprint{hep-th/9611134}.

\end{thebibliography}

\end{document}